\documentclass[12pt]{article}
\usepackage[a4paper, total={7in, 10in}]{geometry}
\usepackage{graphicx}
\usepackage{jhep-mod}
\usepackage{bm}
\usepackage{amssymb,amsmath,amsthm}
\usepackage{mathrsfs}
\usepackage[utf8]{inputenc}
\usepackage{enumerate}
\usepackage{bigints}
\usepackage{xcolor}
\usepackage{appendix}
\usepackage{graphicx}
\usepackage{float}
\usepackage{tikz}
\usepackage{setspace}
\usepackage{cancel}
\usepackage{array}
\usepackage{tabulary}
\usepackage{doi}
\definecolor{purple}{rgb}{1,0,1}
\definecolor{lime}{HTML}{A6CE39} 

\newcommand{\blue}[1]{{\color{blue} #1}}

\definecolor{lime}{HTML}{A6CE39}
\newcommand{\orcidicon}{%
	\begin{tikzpicture}
	\draw[lime, fill=lime] (0,0) 
		circle [radius=0.16] 
		node[white] {{\fontfamily{qag}\selectfont \tiny ID}};
	\draw[white, fill=white] (-0.0625,0.095) 
		circle [radius=0.007];
	\end{tikzpicture}
	\hspace{-5mm}
}
\newcommand\orcidMatt{{\href{https://orcid.org/0000-0003-1088-6485}{\orcidicon}}}

\newcommand{\be}{\begin{equation}}
\newcommand{\ee}{\end{equation}}

\def\sign{\mathrm{{sign}}}
\begin{document}
\newcommand{\arXiv}[1]{arXiv:\href{https://arxiv.org/abs/#1}{\color{blue}#1}}

\title{\vspace{-25pt}\huge{
\blue{Traversable Kaluza--Klein wormholes?}
}}


\author{\Large Christopher Simmonds}
\emailAdd{chris0simmonds@gmail.com}
\author{and \Large Matt Visser\!\orcidMatt$^{\dagger}$}
\emailAdd{matt.visser@sms.vuw.ac.nz}
\affiliation{School of Mathematics and Statistics, Victoria University of Wellington, \\
\null\qquad PO Box 600, Wellington 6140, New Zealand.}
\affiliation{$^\dagger$ Corresponding author.}
\renewcommand{\arXiv}[1]{arXiv:\href{https://arxiv.org/abs/#1}{\color{blue}#1}}
\def\L{{\mathcal{L}}}

\abstract{ \\
Various authors have suggested that Kaluza--Klein variants of traversable wormholes might to some extent ameliorate the defocussing properties (the curvature condition violations, and implied energy condition violations) inherent in positing the existence of a traversable wormhole throat. Unfortunately such a hope is ill-founded. We shall show that in a traditional Kaluza--Klein context the price paid for completely eliminating the  defocussing properties of the wormhole throat is extremely high---to completely eliminate curvature condition violations the 5th dimension has to become truly enormous (formally infinite) in the vicinity of the wormhole throat, in a manner that is fundamentally incompatible with the traditional Kaluza--Klein \emph{ansatz}. 
At best, the extra dimensions allow one to move the curvature condition violations around, they cannot be eliminated except at prohibitive cost.
While traversable Kaluza--Klein wormholes might be interesting for other reasons, it must be emphasized that adding a 5th dimension is not particularly useful in terms of ameliorating violations of the curvature conditions.

\bigskip
\noindent
{\sc Date:} Mon 4 August 2025; Sun 19 October 2025; \LaTeX-ed \today

defocussing; wormhole throat; Kaluza--Klein wormholes.  

\bigskip
\noindent{\sc Published as}: \\
Universe {\bf 11} (2025) 347;
\doi{10.3390/universe1110034} [\arXiv{2508.02824} [gr-qc]] 
\bigskip
\noindent

}

\maketitle
\def\tr{{\mathrm{tr}}}
\def\diag{{\mathrm{diag}}}
\def\cof{{\mathrm{cof}}}
\def\pdet{{\mathrm{pdet}}}
\def\QED{ {\hfill$\Box$\hspace{-25pt}  }}
\def\d{{\mathrm{d}}}
\def\sign{\hbox{sign}}

\parindent0pt
\parskip7pt

\clearpage
\null
\vspace{-75pt}
\section{Introduction}
Both Kaluza--Klein theories and traversable wormholes have a long, complex, and quite subtle history.
The Kaluza--Klein theories date back to the 1920s, with early contributions from Kaluza~\cite{Kaluza, Kaluza:translation} and Klein~\cite{Klein:1926a,Klein:1926b,Ravndal:2013},  partially based on even earlier 1914 work by Nordstr\"om~\cite{Nordstrom:1914,Nordstrom:translation}. 
Over the subsequent century this topic has seen an extremely large number of additional contributions, among which we mention \cite{Modern-KK,Wesson,Overduin:1997, Chodos:1979, Witten:1981a, Witten:1981b, Salam:1981, Freund:1980, Appelquist:1982, Appelquist:1983, Gibbons:1985, Duff:1994}. 

Similarly, wormholes can be traced back to early contributions by Flamm in 1916 ~\cite{Flamm,Flamm:2015}, by Einstein and Rosen in 1935~\cite{Einstein:1935} (the ``Einstein--Rosen bridge''), and by Wheeler in 1955~\cite{Wheeler:1955} (``spacetime foam'').\footnote{See also references~\cite{Hawking:1978, Garay:1998, Carlip:2022}.}

Subsequently, traversable wormholes were popularized by Morris and Thorne some 40 years ago~\cite{Morris-Thorne,MTY} and have now become relatively mainstream (if speculative) objects of scientific interest~\cite{Visser:1989a,Visser:1989b, Hochberg:1990, Frolov:1990, Cramer:1994, Visser:1995, Poisson:1995, Hochberg:1997, Visser:1997, Hochberg:1998a, Hochberg:1998b, Teo:1998, Lemos:2003, Visser:2003, Kar:2004, Lobo:2005, Sushkov:2005, Lobo:2007, Damour:2007, Nakajima:2012, Lobo:2016, Lobo:2017, Boonserm:2018, Simpson:2018, Kuhfittig:2022-HCO, Kuhfittig:2024-minimum}.
A key observation~\cite{Morris-Thorne,MTY} is that wormhole throats, essentially by definition, serve to defocus light rays traversing the throat, and that this defocussing is the harbinger of severely unusual physics --- specifically you need gravity to be repulsive at/near the wormhole throat~\cite{Morris-Thorne,MTY}. This is normally phrased geometrically in terms of violation of the null curvature condition (NCC), or, assuming dynamical equations that are close to the Einstein equations, violations of the null energy condition (NEC)~\cite{Morris-Thorne,MTY}.  

Various ways of (to some extent) ameliorating the energy condition violations have been proposed, sometimes amounting to little more than ``sleight of hand''. 
In the current article we shall adopt a (4+1) dimensional Kaluza--Klein framework, with a view to possibly ameliorating the curvature condition violations --- by modifying the defocussing behaviour at/near the wormhole throat~\cite{Kuhfittig:2025,Kuhfittig:2024,Kuhfittig:2018}.\footnote{And going to even higher dimensionality introduces no essentially new ideas, and does not help.}
 We shall soon see that the price paid for completely eliminating the defocussing behaviour is extremely high --- arguably too high a price to make these particular variants of Kaluza--Klein traversable wormholes physically interesting. 

\clearpage
\section{Area coordinates \emph{versus} proper distance coordinates}
Let us, subject to suitable symmetry restrictions, seek to provide a reasonably general description of ordinary (3+1) dimensional traversable wormholes~\cite{Morris-Thorne,MTY}.
Many authors prefer to use ``area coordinates" (often mis-characterized as  ``Schwarzschild coordinates''\footnote{Schwarzschild himself used somewhat different coordinates~\cite{Schwarzschild:1916}. These area coordinates seem to have first been explicitly introduced by Droste~\cite{Droste:1916,Droste:2002a,Droste:2002b} and Hilbert~\cite{Hilbert:1916} --- they have the very nice feature that the area of a sphere of coordinate ``radius'' $r$ is just $A(r)=4\pi r^2$.}) and write the line-element in the form:
\begin{equation}
\label{E:area-coordinates}
     ds^2=-e^{-2\Phi(r)}dt^2+\frac{dr^2}{1-\frac{b(r)}{r}}+r^2(d\theta^2+\sin^2\theta d\phi^2).
 \end{equation}
In a traversable wormhole context there are good physics and mathematics reasons for instead preferring to use proper distance coordinates. If we adopt ``area coordinates'' as above
then the wormhole throat is located at the root $r_0$ of the equation $b(r)=r$, and the coordinate system presented above at best covers only half the spacetime, from the throat at $r_0$ out to spatial infinity on our side of the wormhole --- at $r=\infty$. We would then need (at a minimum) at least one other coordinate patch of the same type to cover the region from the wormhole throat out to the other spatial infinity on the other side of the wormhole.\footnote{And the region $r<r_0$ would simply be discarded --- it would not be part of the traversable wormhole spacetime.} 

Each of these two coordinate patches would be semi-closed, and they would overlap only at the throat $r_0$ itself, which is sitting right on top of a coordinate singularity.\footnote{Additionally we demand $\Phi(r_0)$ to be finite, so that $e^{-2\Phi(r_0)}$ be non-zero at the throat.}
Note that mathematicians do not like closed or semi-closed coordinate patches, they prefer to work on open coordinate charts. 
So a mathematician would at the very least want to add a third coordinate patch to his/her atlas, one that straddles the throat of the wormhole.\footnote{The situation is even worse for intra-universe wormholes, as opposed to inter-universe wormholes, where one additionally has to worry about how to merge the two spatial infinities into a single asymptotic universe. For simplicity let us consider inter-universe wormholes connecting two separate universes.} 
Physicists, if they are careful, (and we emphasize the need for care and discretion), could get away with using 2 semi-closed coordinate patches. 
Physicists, if they are not careful, could get horribly confused as to the physics of the wormhole throat.\footnote{For a specific  example of what can go wrong if the wormhole throat is sitting right on top of a coordinate singularity, (not a physical singularity), see references~\cite{defective,Feng:2023}.}

\clearpage
Of course you could short-circuit all of these issues, make the mathematicians happy, and get away with a single globally defined coordinate patch if you adopt proper distance coordinates:
\begin{equation}
\label{E:proper-distance-coordinates}
     ds^2=-e^{-2\Phi(\ell)}dt^2+{d\ell^2}+\Sigma(\ell) \; (d\theta^2+\sin^2\theta d\phi^2).
 \end{equation}
 The proper distance $\ell$ coordinate runs over the entire real line $(-\infty,+\infty)$.
The wormhole throat can be chosen (for convenience) to be at $\ell=0$ where $\Sigma(\ell)$ is assumed to have a local minimum. (That is, we want $\Sigma'(\ell=0)=0$ to guarantee a local extremum, and  $\Sigma''(\ell=0)>0$ to guarantee a local minimum). Asymptotically, at large distances from the throat we would want $\Sigma(\ell) \sim \ell^2$. The area of a 2\nobreakdash-sphere at fixed proper distance from the throat is now $A(\ell) = 4\pi \Sigma(\ell)$, which is asymptotic to  $A(\ell) \sim 4\pi \ell^2$ at large distances.
Calculations in this proper distance coordinate system tend to be simpler and easier to interpret; not least because one can now use a single coordinate patch to cover the entire spacetime.

\section{Null geodesic de-focussing at the throat}

\begin{figure}[!h]
    \centering
    \includegraphics[width=0.5\linewidth]{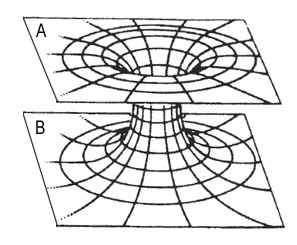}
    \caption{Idealized example of an inter-universe traversable wormhole, connecting two universes ``A'' and ``B''. Note that the wormhole throat acts to de-focus initially incoming converging bundles of light rays, converting them into outgoing diverging bundles of light rays.
    This de-focussing is a key unavoidable characteristic of any geometry one might plausibly wish to call a traversable wormhole.}
    \label{F:wh1}
\end{figure}

A key purely geometrical and intrinsic feature of anything one might want to call a ``traversable wormhole'' is that the wormhole throat should de-focus light rays~\cite{Morris-Thorne}. (An initially converging bundle of light rays impinging on the wormhole throat should become a diverging bundle of light rays, but \emph{without} first passing through a caustic.) See figure \ref{F:wh1} for a schematic representation.

\clearpage
Now ordinary gravitational fields focus light rays, they do not defocus light rays. (Observationally, we see numerous gravitational lenses --- we have not yet seen any evidence for gravitational anti-lensing.\footnote{See for instance reference~\cite{Cramer:1994} for one specific  example of what to look for.}) 
That is, purely on geometrical grounds, (3 + 1) dimensional traversable wormholes must, by definition, violate the so-called null convergence condition (NCC).
\footnote{We will set up a suitable framework for Kaluza--Klein wormholes in a couple of pages.}

Mathematically, after a brief calculation using the Raychaudhuri equation,  this can be expressed in terms of a condition on the Ricci tensor or the Einstein tensor. In the vicinity of the throat there must be some null vectors $\ell^a$, such that in terms of the metric $g_{ab}\ell^a \ell^b=0$, whereas in terms of the spacetime curvature $R_{ab}\ell^a \ell^b<0$, (equivalently $G_{ab}\ell^a \ell^b<0$).\footnote{With the usual sign convention, that the Ricci tensor of a 2-sphere be positive definite.} 

\section{Curvature conditions \emph{versus} energy conditions}

When analyzing wormhole physics, should one use the energy conditions
(constraints on the stress-energy tensor)
or invoke the purely geometrical convergence conditions
(constraints in the Einstein/Ricci tensor)?
This is ultimately as much a sociology of physics question, 
as it is a scientific question. In standard Einstein gravity 
\begin{equation}
G_{ab} = 8\pi G_{Newton} \; T_{ab}.
\end{equation}
So in standard Einstein gravity (energy condition) $\Longleftrightarrow$ (convergence condition). 
But what happens in various modified theories of gravity?

In modified gravity, quite generally the equations of motion are
\begin{equation}
\hbox{some-function} \left( Riemann, metric, T_{ab}, \hbox{other-stuff}\right) = 0.
\end{equation}
Quite often (in fact, almost always) this can be rearranged into the form
\begin{equation}
G_{ab} = \hbox{some-other-function} \left( metric, Weyl, T_{ab}, \hbox{other-stuff}\right).
\end{equation}
When written in this form one has
\begin{equation}
G_{ab} = 8\pi G_{Newton} \; T_{ab}^\mathrm{effective}.
\end{equation}
Then even in many modified gravities one can write
\begin{equation}
 \hbox{(energy condition on $T^\mathrm{effective}_{ab}$)} \Longleftrightarrow\hbox{(convergence condition)}. 
\end{equation}
This has been extremely well known, (if not always well appreciated), for decades.\footnote{For general background, see references~\cite{Curiel:2014, Martin-Moruno:2017, Barcelo:2002}.}

But this is one place where the sociology of physics trumps the actual science.
There is now a truly vast industry churning out endless repetitive and superficial papers on sundry ``wormholes without energy condition violations in modified gravity". But, geometrically, and unavoidably, one is still violating the convergence conditions. 
Still, merely to short circuit some of the chatter,
it might be a good idea (sociologically, if not scientifically)
to focus on the convergence conditions.
Strategically, it is of vital important to usefully communicate with the widest possible cross section of the broader physics community.\footnote{Note, for instance, recent efforts at rephrasing and reanalyzing the singularity theorems as applied to regular black holes in terms of the convergence conditions, as opposed to the energy conditions~\cite{Borissova:2025a,Borissova:2025b}.}

\section{Traditional Kaluza--Klein \emph{ansatz}}

The most basic form of the traditional Kaluza--Klein \emph{ansatz} amounts to adding an  extra dimension to (3+1) Minkowski space, replacing 
\begin{equation}
ds^2= - dt^2 + dx^2+dy^2+dz^2,
\end{equation}
by the (4 + 1) dimensional Lorentzian metric
\begin{equation}
ds^2=  - dt^2 + dx^2+dy^2+dz^2+ L_0^2 \; dw^2; \qquad (w \hbox{ periodic } 2\pi).
\end{equation}
This amounts to adding a compact circle of constant circumference $C_0=2\pi L_0$ to each point in 3+1 dimensional Minkowski space. (This is often called the ``cylinder condition''.) To be compatible with various experiments, $C_0=2\pi L_0$ has to be suitably small. 

Specifically, with a probe of energy $E$ one can operationally resolve distances down to $\Delta x \sim {\hbar c\over E}$, so you want $C_0=2\pi L_0 \ll {\hbar c\over E}$. Given current (and near future) experimental capabilities, (in particular, the maximum beam energy of the LHC~\cite{LHC,LHC2}), you need to set  
 $C_0=2\pi L_0 \ll 10^{-20} \,m$, about $10^{-5} \times \hbox{(nuclear diameter)}$.\footnote{Note that the traditional Kaluza--Klein \emph{ansatz} pre-dates string theory by at least 50 years. Appeals to string theory to justify the Kaluza--Klein \emph{ansatz} are utterly missing the point. }
 When modifying the original Kaluza--Klein \emph{ansatz} for a traversable wormhole context we will want to stay as close as possible to this traditional \emph{ansatz}.\footnote{In particular there is no pressing need (nor advantage) in suppressing the Kaluza--Klein distance scale all the way down to (near) the Planck distance.}

\section{Traversable Kaluza--Klein wormholes}

The basic idea here is to (attempt to) use the 5th dimension to build a traversable wormhole without violating the convergence conditions (and so implicitly satisfy the energy conditions). We shall soon encounter significant difficulties with this idea.

\subsection{Area coordinates}
To go from a (3+1) dimensional traversable wormhole to a (4+1) dimensional traversable wormhole in area coordinates consider the transition from
\begin{equation}
ds^2= - e^{-2\Phi(r)}dt^2 + {dr^2\over1-2m(r)/r} + r^2 \,d\Omega^2 
\end{equation}
to
\begin{equation}
ds^2= - e^{-2\Phi(r)}dt^2 + {dr^2\over1-2m(r)/r} + r^2 \, d\Omega^2 + L(r)^2 \; dw^2; \qquad (w \hbox{ periodic } 2\pi).
\end{equation}
Here $L(r)$ is, (as per the original Kaluza--Klein ansatz), supposed to be ``small enough'' to not be immediately obvious to experimental probes. 
In the Kaluza--Klein literature the quantity $L(r)$ is often referred to as the dilaton field (or sometimes the radion field).
We would want $L(r)\to L_0$ at asymptotically large distances from the wormhole. 
Of course these area coordinates for 4+1 dimensional traversable wormholes inherit all of the technical difficulties, (coordinate singularities at the wormhole throat, the need for multiple coordinate charts), already inherent in the use of area coordinates for 4+1 dimensional traversable wormholes.
We shall again see that it proves easier to work with proper distance coordinates.

\subsection{Proper distance coordinates}

\bigskip
First, recall that in (3+1) dimensions a suitable traversable wormhole \emph{ansatz} is:
\begin{equation}
ds^2= - e^{-2\Phi(\ell)}dt^2 + {d\ell^2} + \Sigma(\ell) \,d\Omega^2. 
\end{equation}
Furthermore, radial null geodesics have cross sectional area
\begin{equation}
A(\ell) = 4\pi \Sigma(\ell).
\end{equation}
The wormhole throat is at a local minimum of $A(\ell)$; and
traversable as long as $e^{-2\Phi(\ell)}>0$. 
Note the NCC is automatically violated.

Now generalize this \emph{ansatz} to (4+1) dimensions:
\begin{equation}
ds^2= - e^{-2\Phi(\ell)}dt^2 + {d\ell^2} + \Sigma(\ell) \, d\Omega^2 + L(\ell)^2 \; dw^2; 
\qquad (w \hbox{ periodic } 2\pi).
\end{equation}
What could possibly go wrong?

Radial null geodesics, (those geodesics with constant $\theta$, $\phi$, and $w$ coordinates), \break because they are now also spread over the entire transverse 5th dimension, $w$,  now have cross sectional \emph{volume}
\begin{equation}
V(\ell) = A(\ell) \times [2\pi L(\ell) ] =  [4\pi \Sigma(\ell)] \times [2\pi L(\ell) ]
= 8\pi^2 \Sigma(\ell) L(\ell).
\end{equation}
It is the interplay between (3 + 1) cross-sectional area $4\pi \Sigma(\ell)$ and transverse 5-th dimensional width $2\pi L(\ell)$ that proponents of possible Kaluza--Klein induced amelioration of  curvature condition violations were (mistakenly) relying on.
We again emphasize that in the Kaluza--Klein literature the quantity $L(\ell)$ is often referred to as the dilaton field (or sometimes the radion field).

Specifically:
If you choose to (somewhat obtusely) still define a wormhole throat by looking for a local minimum of $\Sigma(\ell)$, \emph{and you want to globally satisfy the NCC}, then, (as we shall soon see), $V(\ell)$ cannot have any local minima.
Indeed, \mbox{(4 + 1)} dimensional defocussing (which one is aiming to eliminate) occurs at the local minima of $V(\ell)$, which are not necessarily local minima of $\Sigma(\ell)$.
Eliminating \mbox{(4 + 1)} dimensional defocussing eliminates local minima in  $V(\ell)$, 
which in turn forces the size of the 5th dimension $C(\ell)=2\pi L(\ell)$ to become arbitrarily large (formally infinite) in the vicinity of the traversable wormhole throat.

\section{Spacetime curvature}

To get a clearer and more explicit grip on what is going on, 
let us perform a few specific calculations of appropriate parts of the Ricci tensor.

\subsection{3+1 traversable wormhole in proper distance coordinates}
Take
\begin{equation}
ds^2= - e^{-2\Phi(\ell)}dt^2 + {d\ell^2} + \Sigma(\ell) \,d\Omega^2. 
\end{equation}
Order the coordinates as $x^a=(t,\ell,\theta,\phi)$ and in an orthonormal basis write the radial null vector as $\ell^{\hat a}=(1,1,0,0)$. 

 Then the physically interesting part of the Ricci tensor is
\begin{equation}
    R_{ab} \; \ell^a \ell^b =  R_{\hat a\hat b} \; \ell^{\hat a} \ell^{\hat b} 
    = R_{\hat t\hat t}  + R_{\hat r\hat r}.
\end{equation}
A brief calculation yields
\begin{equation}
    R_{ab} \; \ell^a \ell^b = 
    - {1\over\Sigma(\ell)} {d^2\Sigma(\ell)\over d\ell^2} 
    +  {1\over2\Sigma(\ell)^2} \left(d\Sigma(\ell)\over d\ell\right)^2 
    - {1\over\Sigma(\ell)} \left(d\Sigma(\ell)\over d\ell\right)
    \left(d\Phi(\ell)\over d\ell\right).
\end{equation}
The key point is that at any local extremum of $\Sigma(\ell)$ we have
\begin{equation}
    R_{ab}\;  \ell^a \ell^b \to  - {1\over\Sigma(\ell)} {d^2\Sigma(\ell)\over d\ell^2}. 
\end{equation}
Then if the extremum is a local minimum, the NCC is automatically violated. (Whereas for a local maximum, an anti-throat, the NCC is instead automatically satisfied.)
These (3+1) considerations are utterly standard.
Let us now see how this logic extends to (4+1) Kaluza--Klein wormholes.

\subsection{4+1 Kaluza--Klein wormhole in proper distance coordinates}

We find it  convenient to use 
\begin{equation}
    L(\ell)= {V(\ell)\over8\pi^2\Sigma(\ell)},
\end{equation}
to re-write the spacetime metric as 
\begin{equation}
ds^2= - e^{-2\Phi(\ell)}dt^2 + {d\ell^2} + \Sigma(\ell) \, d\Omega^2 +  {V(\ell)^2\over(8\pi^2)^2\Sigma(\ell)^2} \; dw^2; \qquad (w \hbox{ periodic } 2\pi).
\end{equation}
Order the coordinates as $x^a=(t,\ell,\theta,\phi,w)$, and in an orthonormal basis write the radial null vector ar $\ell^{\hat a}=(1,1,0,0,0)$. Then the physically interesting part of the Ricci tensor is
\begin{equation}
    R_{ab} \; \ell^a \ell^b =  R_{\hat a\hat b} \; \ell^{\hat a} \ell^{\hat b} 
    = R_{\hat t\hat t}  + R_{\hat r\hat r}.
\end{equation}
A brief calculation in this (4+1) context now yields
\begin{eqnarray}
    R_{ab} \; \ell^a \ell^b &=&
     - {1\over V(\ell)} {d^2 V(\ell)\over d\ell^2} 
    -  {3\over2\Sigma(\ell)^2} \left(d\Sigma(\ell)\over d\ell\right)^2 
    \nonumber\\ &&
    - {1\over V(\ell)} \left(d V(\ell)\over d\ell\right) \left(d\Phi(\ell)\over d\ell\right)
    + {2\over \Sigma(\ell) V(\ell)} \left(d V(\ell)\over d\ell\right) \left(d\Sigma(\ell)\over d\ell\right).
\end{eqnarray}
The key point is that at any local extremum of $V(\ell)$ we now have
\begin{equation}
    R_{ab}\; \ell^a \ell^b \to 
     - {1\over V(\ell)} {d^2 V(\ell)\over d\ell^2} 
    -  {3\over2\Sigma(\ell)^2} \left(d\Sigma(\ell)\over d\ell\right)^2 
    \leq - {1\over V(\ell)} {d^2 V(\ell)\over d\ell^2}. 
\end{equation}

\clearpage
Thence to globally satisfy the NCC, $R_{ab} \; \ell^a \ell^b\geq 0$, we must at the very least demand that $ {d^2 V(\ell)\over d\ell^2}\leq 0$ at every extremum of $V(\ell)$. But this means $V(\ell)$ can have no local minima. 
Instead $V(\ell)$ must have some global \emph{maximum} at some $\ell_0$ with
\begin{equation}
    V(\ell) \leq V(\ell_0); \qquad \forall \ell.
\end{equation}

Indeed, rewriting this in terms of $\Sigma(\ell)$ and $L(\ell)$, we see that \emph{globally} we must have
\begin{equation}
\Sigma(\ell_0) L(\ell_0) \geq \Sigma(\ell) L(\ell), \qquad (\forall \ell).
\end{equation}
Thence
\begin{equation}
L(\ell_0) \geq {\Sigma(\ell)\over \Sigma(\ell_0) } \;  L(\ell); \qquad (\forall \ell).
\end{equation}
This observation (assuming you want to globally eliminate curvature condition violations) forces the 5th dimension (in the vicinity of the wormhole throat) to become enormous, macroscopically so,
contrary to the Kaluza--Klein \emph{ansatz}.
In fact, in terms of the original area coordinate $r$,
to avoid NCC violations you need
\begin{equation}
L(\ell_0) \geq {r(\ell)^2\over r(\ell_0)^2 } \;  L(\ell) \qquad (\forall \ell).
\end{equation}
Since at large distances we wanted $L(\ell)\to L_0$, (to respect the traditional Kaluza--Klein \emph{ansatz}), and $r(\ell) \to |\ell|$, (to make the spatial 3-slices asymptotically flat), while we also want $r(\ell_0)= r_0 > 0$, (by definition of a finite size wormhole throat), we see that the ratio $r/r_0$ can be made arbitrarily large. Thence
\begin{equation}
L(\ell_0) \gtrsim {|\ell|^2\over r_0^2 } \;  L_0 \qquad (\forall |\ell| \gg r_0).
\end{equation}
It follows that, in the attempt to avoid NCC violations, you have instead utterly  decompactified the 5th dimension in the vicinity of the wormhole throat.\footnote{Note that no strings were harmed in performing this analysis.}
(Indeed in an infinitely large universe, since $|\ell|$ can be arbitrarily large, the size of the 5th dimension is formally driven to infinity.)
Therefore, the Kaluza--Klein wormhole proposal must be viewed as non-viable, at least insofar as one wishes to avoid all possible NCC violations. Of course if you allow NCC violations,  Kaluza--Klein traversable wormholes are no worse than ``ordinary'' (3+1) dimensional traversable wormholes.

We should emphasize that this global result is very much dependent on eliminating all NCC violations. If you merely want to move the wormhole throat around a little, by adjusting $\Sigma(\ell)$ and possibly $L(\ell)$ independently, that can certainly be easily achieved. 

More subtly, as we have seen, the NCC violations are associated with local minima of $V(\ell)$, which (in the absence of $Z_2$ symmetry under interchange of the two universes) are not necessarily local minima of $\Sigma(\ell)$, or local extrema of $L(\ell)$. 
So if one rather obtusely tries to define the wormhole throat by looking for a local minima of $\Sigma(\ell)$, then you are likely looking in the wrong place.

\section{Discussion}

In summary, while  Kaluza--Klein variants of traversable wormholes might be of some interest in their own right, it must be emphasized that adding a 5th dimension will do nothing to ameliorate the convergence condition violations, at least not without the unacceptably high price of decompactifying the 5th dimension in the vicinity of the wormhole throat.

Of course there are many other (quite different and distinct) reasons for  potentially being interested in both Kaluza--Klein and related multi-dimensional traversable wormholes, reasons that would be quite unaffected by the arguments herein. 
One could either accept the NCC violations and work with them, or alternatively adopt even more radical frameworks. 
In particular, membrane-based wormhole models would provide a completely different way of exploring higher dimensions~\cite{Visser:1985, Rubakov:1983, Randall:1999, Bronnikov:2002, Lobo:2007b}, and require a rather different analysis.


\bigskip
\bigskip
\hrule\hrule\hrule

\clearpage
\bigskip
\hrule\hrule\hrule
\addtocontents{toc}{\bigskip\hrule}

\setcounter{secnumdepth}{0}
\section[\hspace{14pt}  References]{}
%

\end{document}